\documentclass[journal]{IEEEtran}
\usepackage{amsmath}
\usepackage{amsfonts}

\newcommand{\RNum}[1]{\uppercase\expandafter{\romannumeral #1\relax}}
\usepackage{booktabs}
\usepackage{longtable}
\usepackage{tabularx}
\usepackage{siunitx}
\ifCLASSINFOpdf
\usepackage[pdftex]{graphicx}
\else
\fi
\begin{document}
%
\title{Benchmark of Deep Learning-based Imaging PPG in Automotive Domain}
%
%
%


\author{\IEEEauthorblockN{Yuqi~Tu\IEEEauthorrefmark{1,}\IEEEauthorrefmark{2},
        Shakith Fernando\IEEEauthorrefmark{2},
        Mark van Gastel\IEEEauthorrefmark{2} 
}

\IEEEauthorblockA{\IEEEauthorrefmark{1}Eindhoven University of Technology, Eindhoven}

\IEEEauthorblockA{\IEEEauthorrefmark{2}Philips Intellectual Property \& Standards, Eindhoven}}
\maketitle

\begin{abstract}
Imaging photoplethysmography (iPPG) can be used for heart rate monitoring during driving, which is expected to reduce traffic accidents by continuously assessing drivers’ physical condition. Deep learning-based iPPG methods using near-infrared (NIR) cameras have recently gained attention as a promising approach. To help understand the challenges in applying iPPG in automotive, we provide a benchmark of a NIR-based method using a deep learning model by evaluating its performance on MR-NIRP Car dataset. Experiment results show that the average mean absolute error (MAE) is 7.5 bpm and 16.6 bpm under drivers’ heads keeping still or having small motion, respectively. These findings suggest that while the method shows promise, further improvements are needed to make it reliable for real-world driving conditions.
\end{abstract}


%
\IEEEpeerreviewmaketitle

\section{Introduction}
%
%
%
%
\IEEEPARstart{I}{n} future vision of mobility safety \cite{euroncap}, continuous monitoring of drivers' vital signs like heart rate is essential. To achieve this goal, Imaging photoplethysmography (iPPG) is applied. iPPG estimates drivers' pulse signal in a non-contact manner by detecting subtle color changes in facial skin, captured by a camera installed in the vehicle. However, varying lighting conditions while driving significantly affect the results of RGB three-channel cameras largely. In contrast, a NIR camera with a specific optical density bandpass filter is more robust to light variation during driving\cite{NowaraDriving}, offering a promising alternative. Recent studies\cite{9506663}\cite{yu2019remote}\cite{10301524} have utilized NIR cameras in combination with using deep learning model to explore accurate drivers' heart rate estimation methods. To help understand the advances and challenges in drivers' heart rate estimation, we provide a benchmark of a representative deep learning-based approach\cite{9506663} in this report on the MR-NIRP Car dataset\cite{NowaraDriving}.

\section{Related Work}
For deep learning-based iPPG using NIR camera, \cite{9506663} proposed a novel U-net architecture\cite{ronneberger2015u}. Other approaches estimate heart rate (HR) using different techniques. In \cite{yu2019remote}, an end-to-end method based on a 3D Convolutional Neural Network (3DCNN) and a Recurrent Neural Network (RNN) with images as input is proposed. In \cite{10301524}, Li-Wen \textit{et al.} proposed an approach training two deep learning modules, one encoder-decoder CNN-based model to estimate pulse waveform and a CNN-based HR estimator to derive HR from the estimated waveform. \cite{gideon2021way} modifies the 3DCNN from \cite{yu2019remote} and applies the maximum cross correlation (MCC) loss function to address the time offset existing between the estimated waveform and reference label. 

For benchmark on iPPG of drivers, \cite{10095078} evaluates various deep learning-based methods excluding \cite{9506663}. According to the results in \cite{10095078} and \cite{9506663}, \cite{9506663} delivers a more accurate estimation. However, \cite{9506663} only evaluates its performance under a few conditions by limited metrics. To provide a more comprehensive evaluation of the method in [3], this report presents a benchmark demonstrating the method's performance under various conditions and using a broader set of metrics. In addition to the benchmark, we propose several variant methods and evaluate them, aiming to overcome the limitations we found in the approach of \cite{9506663} and to explore alternative possibilities.

\section{Methodology}
Based on \cite{9506663}, this section introduces the deep learning approach used for benchmarking. The approach consists mainly of two modules, a time series extraction module and deep neural architecture for PPG estimation. Figure.1 demonstrates the overall procedure. Details are explained as follows.
\begin{figure}[!t]
\centering
\includegraphics[width=3.6in]{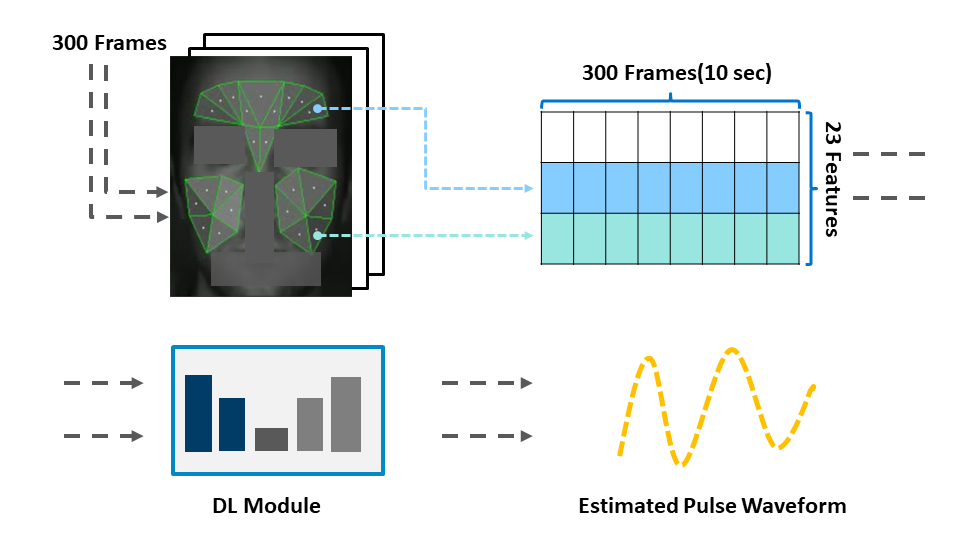}
\caption{Graphical representation of overall procedure of the deep learning-based NIR iPPG method proposed in \cite{9506663}}
\label{fig_sim}
\end{figure}

\subsection{Time series extraction module}
This module serves for \romannumeral 1) feature extraction and \romannumeral 2) processing. Features are extracted from face videos. Different from \cite{9506663}, for each frame in the video, we localize 478 facial landmarks using MediaPipe\cite{Mediapipe} and select 23 regions of interest (ROIs) (where contain strongest PPG signals \cite{kumar2015distanceppg}) instead of 48. 

To begin, a moving window of 300 frames (equivalent to 10 seconds at 30 fps) is applied to the frame series. To facilitate deep learning training, the features within each window are normalized to have a zero mean and constrained to a smaller scale. Each window is used to predict a piece of pulse waveform of the same period, with a stride of 10 frames. The heart rate (HR) is consequently determined by identifying the frequency component of the pulse waveform with the highest magnitude, updating the HR approximately every 0.3 seconds. The ground-truth labels are also windowed with the same size as the frame series. Given the natural cardiac frequency range, the labels are filtered using a band-pass filter between 45 and 180 beats per minute (bpm).

\subsection{Deep neural architecture for PPG estimation}
U-net architecture\cite{ronneberger2015u} is applied in this module to do estimation. Figure.2 demonstrates the visualization of the architecture. The deep neural network extracts the 300-length, 23-dimension time series, as described in section \RNum{3}. \textit{A}.

This U-net model consists primarily of the downsampling section, the upsampling section, and the skip connection. The downsampling section is responsible for feature extraction and the upsampling section restores the resolution lost during downsampling. The skip connections link the downsampling and upsampling modules, aiding in the training process. In \cite{9506663}, it introduces a novel gated recurrent unit (GRU)-based, but such technique is not contained in this report due to its limited upgrade on performance.


As the frequency character is crucial for HR estimation, the goal for training can be interpreted as making the estimated time series linearly related to the ground-truth label. So, same as \cite{9506663}, the deep learning module trains to maximize the Pearson correlation between the estimated waveform and ground-truth label.


\begin{figure}[!t]
\centering
\includegraphics[width=3.6in]{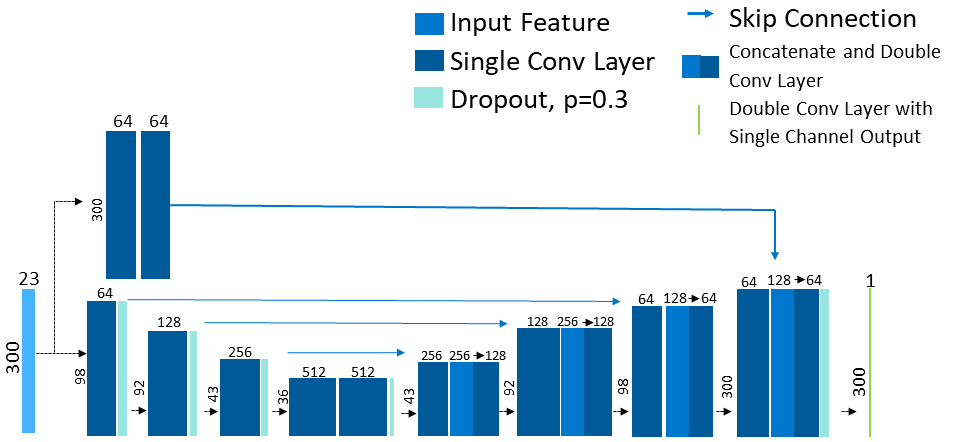}
\caption{U-net model-based Deep learning module used for pulse waveform estimation. Input for the model is 300-length time series with 23 dimensions and it outputs a 300-length, single-channel time series.}
\label{fig_sim}
\end{figure}

\section{Experiment}
This section is composed of two parts. Experiment details will be elaborated in the experiment setup, including more information of the dataset, deep learning training configuration, and data augmentation. Experimental results demonstrate the experiment results and analysis.
Following that are the complete results including motion scenarios. Analysis based on results is contained as well
\subsubsection{Experiment setup}
\textbf{Dataset:} The MERL-Rice Near-Infrared Pulse (MR-NIRP) Car Dataset \cite{NowaraDriving} is used for training and verification. This dataset contains different drivers' face video recorded with an NIR camera. The video frame rate is 30fps and is filtered with a 940 \(\pm \) 5 nm bandpass filter. There are 18 subjects in total and the video is recorded separately during two scenarios, either \textit{driving} (city driving) or \textit{Garage} (parked with engine running). Within each scenario, videos are separated in terms of head motion condition, this report contains test results of two motion levels, \textit{Still} (driver's head keeps still), \textit{small motion}. The ground-truth pulse waveform is collected by g a CMS 50D+ finger pulse oximeter recording at 60 fps and then downsampled to 30 fps to match face videos.

\textbf{Training and test protocols:} The train-test
protocol we apply, as the same in \cite{9506663}, is leave-one-subject-out cross-validation. We trained the model for 8
epochs with batch size 96. Adam optimizer\cite{kingma2014adam} is used with a learning rate $1.0\times 10^{-4}$ reducing after each epoch by a factor of 0.05. 

\textbf{Data Augmentation:} Dataset contains HR ranging from 40 to 110 bpm, but it is unbalanced across HR. Most subjects' HR falls in the range [50,70] bpm. Considering the train-test protocol applied, once the extreme samples are put into the test dataset, performance suffers from the training set lacking such kind of data. To address such a problem, our implementation applies a data augmentation technique based on\cite{9506663}. For both input time series and label, they are resampled with linear resampling rates 1+$r$ and 1-$r$, where $r$ is randomly chosen from the range [0.2,0.6]. This technique is also expected to compensate for the gaps in the distribution of subjects' HR.

\textbf{Metrics:} The performance is evaluated in terms of 3 metrics. The first metric is PTE6 (percent of the time the error is less than 6 bpm). If the absolute error between estimation and ground-truth HR is less than 6, the estimation is regarded as a successful estimation. The threshold 6 is the expected frequency resolution of a 10-second window. The second metric is root-mean-squared error (RMSE), which is defined by calculating root-mean-squared error of each 10-second window and averaged over the test sequence. The third metric is mean-absolute error (MAE), which is defined by calculating the mean-absolute error of each 10-second window and averaging over the test sequence.

\subsubsection{Experiment results} 

\textbf{Comparison Results: }Table \RNum{1}. compares our implementation (labeled with Benchmark) with results shown in \cite{9506663} (labeled with TURNIP). \cite{9506663} demonstrates results of both scenarios under \textit{Still} condition, measured in PTE6 and RMSE. DA stands for data augmentation. For \textit{Garage Still} without data augmentation, the performance of our implementation is comparable to TURNIP's in terms of PTE6 but RMSE is 15\% higher. With augmentation, TURNIP sees a larger upgrade in both metrics. For \textit{Driving Still}, overall performance gets substantially worse and the performance of our implementation is worse than TURNIP's. Our implementation doesn't benefit from data augmentation in terms of the mean of both metrics.
\begin{table}[htbp]
\caption{ HR estimation results (mean $\pm$ std)}
\label{my-label}
\centering
\scriptsize 
\begin{tabularx}{\columnwidth}{l *{4}{>{\centering\arraybackslash}X}}
    \toprule
    & \multicolumn{2}{c}{Driving Still} & \multicolumn{2}{c}{Garage Still} \\
    \cmidrule(lr){2-3} \cmidrule(lr){4-5}
    & PTE6 (\%) & RMSE (bpm) & PTE6 (\%)  & RMSE (bpm)\\
    \midrule
    TURNIP(No DA) & 61.9 $\pm$  22.3 & 10.7 $\pm$  5.9 &  81.9 $\pm$  31.0 & 5.9 $\pm$  8.9 \\
    Benchmark(No DA) & 56.1 $\pm$ 24.2 & 11.9 $\pm$ 6.3 & 81.2 $\pm$ 30.0 & 6.8 $\pm$ 9.4 \\   
    TURNIP & 65.1 $\pm$ 13.9 & 11.4 $\pm$ 4.1 & 89.7 $\pm$ 15.7 & 4.6 $\pm$ 4.8 \\
    Benchmark & 54.2 $\pm$ 15.0 & 17.2 $\pm$ 4.8 & 84.5 $\pm$ 19.2 & 10.1 $\pm$ 8.4 \\ 
    \bottomrule
\end{tabularx}
\end{table}

In the video of \textit{Driving Still}, it's common to see outer illumination's impact on faces' ROIs even though NIR camera is used. During these moments, feature extraction gets noisy and both training and test are affected, which leads to the performance downgrade.

Two reasons might explain why \cite{9506663} outperforms our implementation, \romannumeral 1) \cite{9506663} uses 48-dimensional time series, which means they select 48 facial regions instead of 23, regions are sliced into smaller ones. In this way, subtle changes could be better observed so that the estimation could be more accurate. \romannumeral 2) Our implementation doesn't contain the adapted skip connection proposed in \cite{9506663}. 

Figure.3 shows two examples of HR estimation. Figure.3 (A) illustrates an accurate estimation, while Figure.3 (B) presents an example with low estimation accuracy from another subject. According to the ground-truth labels, the subject’s HR in Figure.3 (B) remains mostly between 40 and 50 bpm during the recording, but the estimated HR is significantly higher. This discrepancy is likely due to the lack of low HR data in the training set. Data augmentation effectively improves estimation on extreme subjects like this one and the impact of it with more details is shown later. 

\begin{figure}[!t]
\centering
\includegraphics[width=3.6in]{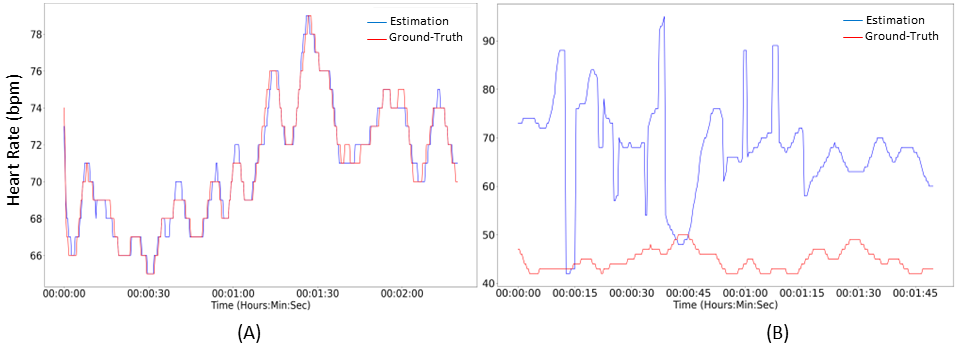}
\caption{Visualization examples of heart rate estimation}
\label{fig_sim}
\end{figure}

Figure.4 shows two examples of estimation results in the frequency domain under \textit{Garage Still}. For the spectrogram, it also updates at the same pace, 0.3 seconds. The spectrum for each window is expected to concentrate around a peak, as the peak reflects the heart rate (HR) for that period, which is assumed to remain stable. It can be observed that in Figure.4 (A), power concentrates on the peak for most of the time while for Figure.4 (B), the spectrum doesn't focus as in (A), indicating the model might have difficulty in providing a reliable estimated waveform.
\begin{figure}[!t]
\centering
\includegraphics[width=3.6in]{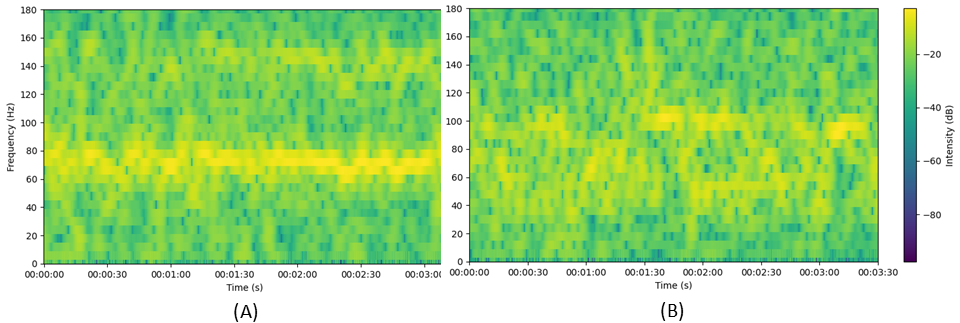}
\caption{Visualization examples of the spectrogram of estimated pulse waveform}
\label{fig_sim}
\end{figure}

Figure.5 demonstrates the impact of data augmentation on \textit{Still} condition of \textit{Garage} scenario. There are two extreme cases with high HR range and low HR range, respectively, while others' HR mostly falls in the normal range. As the figure indicates, with data augmentation, the estimation on extreme cases improves largely, for two subjects with almost 0 accuracy without DA, it goes up to 40\% and 93\%, respectively. Although PTE6 in other subjects sees a slight decrease, overall PTE6 increases by 4.1\%. 
\begin{figure}[!t]
\centering
\includegraphics[width=3.6in]{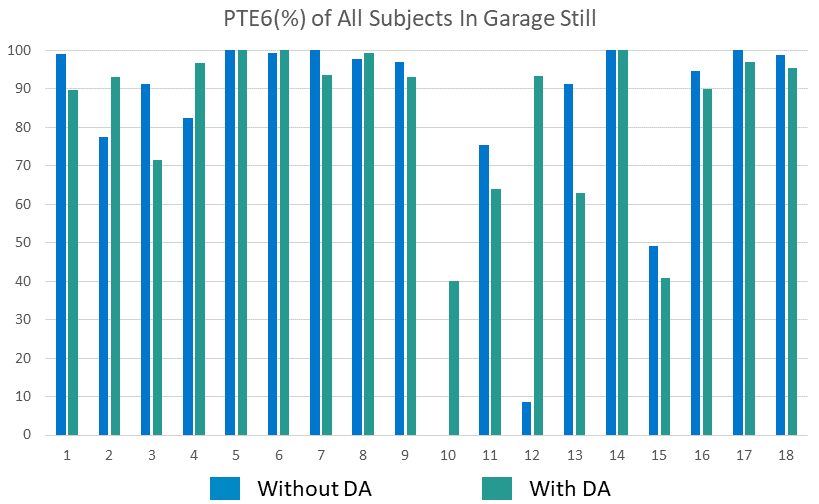}
\caption{Subject-wise performance comparison demonstrating impact of data augmentation on \textit{Still} condition of \textit{Garage} scenario. DA stands for data augmentation}
\label{fig_sim}
\end{figure}

We also tried to investigate why data augmentation doesn't improve performance on \textit{Still} condition of \textit{Driving} scenario. Figure.6 shows a subject-wise comparison, measured in terms of metric PTE6. Although estimation on subjects with extreme HR range improves as well, downgrade on other subjects becomes more obvious and leads to a decrease in average PTE6. This downgrade could be attributed to limitations in the current data augmentation process. Since augmented data often differs from true features, generated data with HR values within the normal range (around 50-70 bpm) may conflict with actual data and mislead the model. Additionally, the resampling operation is not robust enough to handle noise in the original features. As a result, further improvements in data augmentation are still needed.

Subject-wise results reveal that while the overall average estimation accuracy of the implemented approach appears acceptable, the accuracy varies significantly between individual subjects. This indicates that, with the current dataset, the implemented approach still requires improvement to ensure reliability and robustness.

\begin{figure}[!t]
\centering
\includegraphics[width=3.6in]{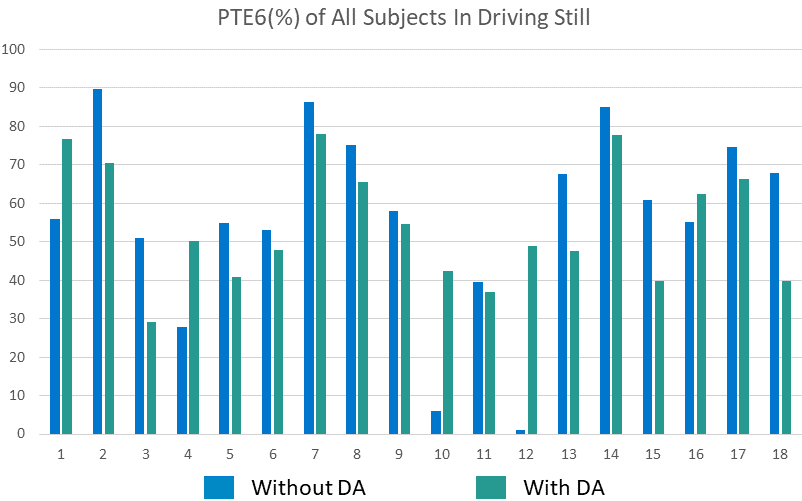}
\caption{Subject-wise performance comparison demonstrating impact of data augmentation on \textit{Still} condition of \textit{Driving} scenario. DA stands for data augmentation}
\label{fig_sim}
\end{figure}

\textbf{Comprehensive Evaluation:}
To provide a more comprehensive benchmark, we test the performance on 2 scenarios under both \textit{Still} and \textit{Driving} conditions. In addition, metric MAE is included, which serves to evaluate performance more comprehensively. Table \RNum{3}. demonstrates the results.  In terms of MAE, \textit{Motion} condition increases MAE by more than 6 bpm for the same scenario. It's apparent that \textit{Motion} of drivers poses a large challenge for estimation. From the video under \textit{Motion}, we can tell that tracking ROIs becomes difficult and feature extraction is then less accurate. Again, data augmentation applied doesn't benefit in \textit{Motion} condition. 
\begin{table}[htbp]
    \centering
    \caption{HR estimation results}
    \label{tab:results}
    \sisetup{
        table-format=2.1(2.1),
        table-number-alignment=center,
        separate-uncertainty=true, 
        tight-spacing=true,         
        }
    \begin{tabular}{lSSS}
        \toprule
        {Condition} & {PTE6 (\%)} & {RMSE (bpm)} & {MAE(bpm)} \\
        \midrule
        Garage\_Still      & 81.2 \pm 30.0 & 6.7 \pm 9.4  & 5.1 \pm 9.1 \\
        Garage\_Still\_DA  & 84.5 \pm 19.2 & 10.1 \pm 8.4 & 5.0 \pm 5.7 \\
        Driving\_Still     & 56.1 \pm 24.2 & 11.9 \pm 6.3 & 8.7 \pm 6.7 \\
        Driving\_Still\_DA & 54.2 \pm 15.0 & 17.2 \pm 4.8 & 11.1 \pm 4.4 \\
        Garage\_Motion     & 39.3 \pm 20.9 & 15.7 \pm 7.0 & 12.0 \pm 6.9 \\
        Garage\_Motion\_DA & 26.6 \pm 14.0 & 23.8 \pm 6.9 & 18.6 \pm 6.7 \\
        Driving\_Motion    & 26.8 \pm 10.6 & 18.9 \pm 6.9 & 14.8 \pm 6.3 \\
        Driving\_Motion\_DA & 19.5 \pm 8.8 & 25.9 \pm 6.6 & 20.8 \pm 6.8 \\
        \bottomrule
    \end{tabular}
\end{table}

\section{Variant Methods}
Taking \cite{9506663} as the backbone, we implemented several variants to improve the performance. Due to the time limit, we focus on \textit{Still} condition in this report.
Details are as follows. 

\textbf{Window Shift:} Considering ground-truth labels and features used to estimate are extracted from different body parts (finger and face), there might be a time offset between these two signals \cite{gideon2021way}.  To address this mismatch, we introduce a window shift technique. The goal of this method is to align the most linearly related sections of the label and the estimated waveform. Still using Pearson correlation, we apply the window shift and modify the loss function from \cite{9506663} in two ways as explained below.


\RNum{1}) After obtaining the estimated waveform in time series form, we truncate both the label and the estimated time series to the same length, denoted by $x$, where $x$ iterates from 0 to 15 (other upper limits were also tested, yielding similar performance). However, the truncation starts from opposite ends for the two series, responsible for canceling out the offset. We then assess the linear relation by calculating the Pearson correlation between the label and the estimation for each value of $x$. When iteration and computation are done, the loss is then defined by the maximum correlation. This method is called WS-1.

\RNum{2}) Still based on window shift, for this method, we enlarge the input series of the deep learning model to 350 frames, so the estimation series length is then also 350. However, the label series length remains the same. To match the label size, a moving 300-length window at stride 1 will iterate within the estimation series, similar to  WS-1, and then it will select a 300-length continuous part of the estimation series which mostly linearly relates with the label series by comparing Pearson correlation and take the correlation value as loss. This method is proposed out of that method WS-1 might lead to inefficiency in dataset usage due to every window being truncated to a certain size. In addition, this method ensures that the HR is still estimated based on the same-length window as \cite{9506663} all the time. We call this method as WS-2. 

\textbf{Fully Deep Learning-based (FDL) Estimation:} We also explored a method that estimates HR completely by deep learning module instead of doing FFT following estimating pulse waveform. In the previous implementation, we found that loss is not always proportional to HR estimation performance, which can be explained by that correlation doesn't 100\% inform the frequency characters of the waveform. Hence, training with this loss function might not be the most efficient option. To explore other possibilities, we tried training a deep learning module to directly estimate HR using another loss function. We first implemented this thought by simply adding a downsampling module composed of three fully connected layers after the U-net model described in section \RNum{3}.\textit{B} and the loss function now is the mean square error (MSE). However the experiment result is much worse than what is shown in Table \RNum{2}. Considering the U-net model applied in \RNum{3}.\textit{B} mainly aims to provide a high-resolution waveform and that the added downsampling module is not fully investigated, such performance is foreseeable. \cite{10301524} proposed a deep learning architecture estimating HR whose input is pulse waveform, it's called CNN-Based HR Estimator and claims it's able to eliminate noise contained in the estimated waveform. To further investigate the performance of a fully Deep learning-based estimation method, we combine the HR estimator from \cite{10301524} and the  U-net model in section \RNum{3}.\textit{B}. Since the output of the deep learning module is now HR,  we apply MSE as loss function. Specifically, the loss is defined by MSE between a batch of estimated HR and a batch of label HR while label HR comes from the FFT result of the label pulse waveform. We tried training the U-net model and HR estimator together and only trained the HR estimator taking input from the trained U-net model. Only the latter method can provide comparable performance relative to the original implementation of \cite{9506663} and it is shown in the next paragraph. We call this method FDL.

\textbf{Experimental Results of Variant Methods:} Table \RNum{3} compares three variant methods and original implementation (labeled with Benchmark). Data augmentation is not applied now due to it's its failure in previous implementation and the fact that it's a temporal solution to deal with a lack of data.

\begin{table}[htbp]
\caption{ HR estimation results}
\label{my-label}
\centering
\scriptsize 
\begin{tabularx}{\columnwidth}{l *{6}{>{\centering\arraybackslash}X}}
    \toprule
    & \multicolumn{3}{c}{Driving Still} & \multicolumn{3}{c}{Garage Still} \\
    \cmidrule(lr){2-4} \cmidrule(lr){5-7}
    & PTE6 (\%) & RMSE (bpm)& MAE(bpm) & PTE6 (\%) & RMSE (bpm)& MAE(bpm) \\
    \midrule
    Benchmark & 56.1 $\pm$ 24.2 & 11.9 $\pm$ 6.3 & 8.7 $\pm$ 6.7  & 81.2 $\pm$ 30.0 & 6.8 $\pm$ 9.4 & 5.1 $\pm$ 9.1 \\  
    WS-1 & 58.7 $\pm$ 25.9 & 10.0 $\pm$ 6.2 & 7.6 $\pm$ 6.7 & 81.3 $\pm$ 29.3 & 6.8 $\pm$ 9.2 & 5.1 $\pm$ 9.0 \\ 
    WS-2 & 59.1 $\pm$ 26.3 & 11.0 $\pm$ 6.3 & 7.9 $\pm$ 6.7 & 79.5 $\pm$ 30.7 & 6.7 $\pm$ 9.0 &  5.1 $\pm$ 8.8 \\ 
    FDL & 53.2 $\pm$ 25.4 & 10.5 $\pm$ 7.0 & 8.5 $\pm$ 7.2 & 76.1 $\pm$ 32.9 & 6.8 $\pm$ 8.7 &  5.8 $\pm$ 8.7 \\ 
    
    \bottomrule
\end{tabularx}
\end{table}

As Table \RNum{3} implies, under \textit{Driving Still}, WS-1 and WS-2 provide better performance in terms of all three metrics, especially MAE. WS-1 and WS-2 reduce MAE by 12.6\% and 9.2\%, respectively. FDL also sees slight improvement in terms of RMSE and MAE, but PTE6 is 5.2\% lower than the original implementation. In \textit{Garage Still}, all methods provide comparable performance while the original implementation and WS-1 are slightly better in terms of PTE6. 

Proposed variant methods also have limitations. For window shift-based methods, both WS-1 and WS-2 randomly initialize the parameters of the deep learning module. So it can be expected that time offset still exists, to be more specific, in that the estimated waveform of the very beginning stage is generated by initialization, selected label-estimation pair with the highest correlation is less meaningful and potentially misleading. This could cause the model to train in the wrong direction. For FDL, performance in terms of PTE6 under both conditions is noticeably lower than the benchmark, although the other two metrics remain comparable. This phenomenon can be explained by that large overlap between waveforms of neighboring windows leading to slow variation in FFT results, however, for the deep learning module of FDL, subtle changes in input could lead to large fluctuation on the estimation of HR, even if the average remains consistent, greater fluctuation increases the likelihood of more samples crossing the threshold set by PTE6.

\section{Conclusion}
In this report, we provide a benchmark of deep learning-based imaging PPG during driving. We implement a U-net model-based method, propose variant methods, and test them under different situations. Analysis of the current method and dataset are also included. While estimation under certain conditions shows that deep learning-based imaging PPG is promising, performance in practical situations (driving on roads, drivers' heads moving, etc) still needs to improve.


%



\section*{Acknowledgment}

This assignment is supported by the Philips Study (SEF-2024-300365).We express our sincere gratitude to Inge Lamberts, for her support to the study request. We also express our sincere gratitude to all the drivers who willingly participated in the dataset collection.

\ifCLASSOPTIONcaptionsoff
  \newpage
\fi



%


\bibliographystyle{IEEEtran}  
\bibliography{bibtex/bib/references}

%








\end{document}